\documentclass[prb,aps,twocolumn,showpacs,superscriptaddress,floatfix]{revtex4}
\usepackage{epsfig}
\usepackage{amsmath}
\usepackage{amssymb}
\usepackage{amsfonts}
\usepackage{mathptmx}
\usepackage{eucal}
\usepackage{bm}

\DeclareMathOperator{\tr}{\mathop{\mathrm{Tr}}}

\DeclareMathOperator{\re}{\mathop{\mathrm{Re}}}
\DeclareMathOperator{\im}{\mathop{\mathrm{Im}}}
\DeclareMathOperator{\arctanh}{arctanh}

\newcommand{\Eq}[1]{Eq.~(\ref{#1})}
\newcommand{\Eqs}[1]{Eqs.~(\ref{#1})}
\newcommand{\W}{\widetilde{W}}
\begin{document}

\title{Electron cooling by diffusive normal metal - superconductor tunnel junctions}
\author{A.~S.~Vasenko}
\affiliation{LPMMC, Universit\'{e} Joseph Fourier and CNRS, 25 Avenue des
Martyrs, BP 166, 38042 Grenoble, France}
\author{E.~V.~Bezuglyi}
\affiliation{Institute for Low Temperature Physics and Engineering, Kharkov
61103, Ukraine}
\author{H.~Courtois}
\affiliation{Institut N\'{e}el, CNRS and Universit\'{e} Joseph Fourier, 25
Avenue des Martyrs, BP 166, 38042 Grenoble, France}
\author{F.~W.~J.~Hekking}
\affiliation{LPMMC, Universit\'{e} Joseph Fourier and CNRS, 25 Avenue des
Martyrs, BP 166, 38042 Grenoble, France}
\date{\today}

\begin{abstract}
We investigate heat and charge transport in NN$'$IS tunnel junctions in the
diffusive limit. Here N and S are massive normal and superconducting electrodes
(reservoirs), N$'$ is a normal metal strip, and I is an insulator. The flow of
electric current in such structures at subgap bias is accompanied by heat
transfer from the normal metal into the superconductor, which enables
refrigeration of electrons in the normal metal. We show that the two-particle
current due to Andreev reflection generates Joule heating, which is deposited in
the N electrode and dominates over the single-particle cooling at low enough
temperatures. This results in the existence of a limiting temperature for
refrigeration. We consider different geometries of the contact: one-dimensional
and planar, which is commonly used in the experiments. We also discuss the
applicability of our results to a double-barrier SINIS microcooler.
\end{abstract}

\maketitle

%%%%%%%%%%%%%%%%%%%%%%%%%%%%%%%%%%%%%%%%%%%%%%%%%%%%%%%%%%%%%%%%%%%%%%%%%%%%
%%%%%%%%%%%%%%%%%%%%%%%%%%%%%%%%%%%%%%%%%%%%%%%%%%%%%%%%%%%%%%%%%%%%%%%%%%

\section{Introduction}

The flow of electric current in NIS (Normal metal - Insula\-tor -
Superconductor) tunnel junctions is accompanied by heat transfer from the
normal metal into the supercon\-duc\-tor.\cite{Nahum, Giazotto2006, Heikkila} This
phenomenon arises due to selective tunneling of high-energy quasiparticles out
of the normal metal which is induced by the superconducting energy gap. It is
similar to the Peltier effect in metal-semiconductor contacts\cite{Semicond}
and enables refrigeration of electrons in the normal metal. The heat current
out of the normal metal (also referred to as ``cooling power'') is maximal at a
voltage bias just below the energy gap, $eV \lesssim \Delta$. For $eV \gtrsim
\Delta$ both the current $I$ through the junction and the Joule heating power
$IV$ strongly increase, rendering the cooling power negative.

A micrometer-sized refrigerator, based on a NIS tunnel junction, has been first
fabricated by Na\-hum {\it et al.} \cite{Nahum} The authors used a single NIS
junction in order to cool a small normal metal strip. Later Leivo {\it et
al}\cite{LPA} noticed that the cooling power is an even function of the applied
voltage, and fabricated a refrigerator with two NIS junctions arranged in a
symmetric series configuration (SINIS). This results in reduction of the
electron temperature from 300 mK to about 100 mK, offering perspectives for the
use of NIS junctions for on-chip cooling of nano-sized systems, like
high-sensitive detectors and quantum devices.\cite{on-chip} To enhance the
performance of NIS microcoolers, it is important to understand possible
limitations of the NIS refrigeration.

Serious limitations of the cooling effect arise from the fact that nonequilibrium
quasiparticles injected into the superconducting electrode accumulate near
the tunnel interface.\cite{Sukumar2, VH} The consequences are the backtunneling
of hot quasiparticles to the normal metal\cite{Jug, VH}, the emission of phonons
(by the recombination of nonequilibrium quasiparticles into Cooper pairs) that partially
penetrate the normal metal,\cite{Sukumar2, Jug} and the overheating of the superconducting
electrode.\cite{Sukumar2} All these effects reduce the efficiency of NIS refrigerators.
This problem can be solved by imposing a local thermal equilibrium in the superconducting
electrode.\cite{VH} So called quasiparticle traps,\cite{traps,GV} made of an additional
normal metal layer covering the superconducting electrode, remove hot quasiparticles
from the superconductor and are thus beneficial in this respect.

However, there is a fundamental limitation for NIS microcoolers. It arises from
the intrinsic multiparticle nature of current transport in NIS junctions which
is governed not only by single-particle tunneling but also by two-particle
(Andreev) tunneling. The single-par\-ticle current and the associated heat
current are due to quasiparticles with energies $E > \Delta$ (compared to the
Fermi level). At very low temperatures, single-par\-ticle processes are
exponentially suppressed in the subgap voltage region $eV < \Delta$, and the
charge is mainly transferred by means of Andreev reflection of quasiparticles
with energies $E < \Delta$.\cite{Andreev, S-J} The Andreev current $I_A$ does
not transfer heat through the NS interface but rather generates the Joule
heating $I_A V$ which is deposited in the normal metal electrode\cite{Sukumar1}
and dominates single-particle cooling at low enough temperatures. Thus the
interplay between the single-par\-ticle tunneling and Andreev reflection sets a
limiting temperature for the refrigeration.

The role of the Andreev current in the electron refrigeration has been first
theoretically analyzed by Bardas and Averin for the simplest model of the NIS
microcooler -- a one-dimensional constriction between the N and S
reser\-voirs,\cite{BA} assuming the constriction length to be much shorter than
the coherence length. In experiment, the importance of Andreev processes in NIS
microcoolers was first demonstrated by Rajauria {\it et al},\cite{Sukumar1} by
using the theoretical estimations of the Andreev current,\cite{Hekking}
obtained within the tunnel Hamiltonian technique, for interpretation of the
experimental data. In this paper we present a quantitative analysis of heat
transport in diffusive NIS tunnel junctions based on the solution of
microscopic equations of nonequilibrium superconductivity.\cite{LOnoneq} We
consider the general case of arbitrary length of the normal wire, as well as of
different possible geometries of the junction: one-dimensional (1D) junctions
and planar junctions with overlapping thin-film electrodes, commonly used in
experiments. \cite{Sukumar1} We also discuss the applicability of our results
to a double-barrier SINIS microcooler.

The paper is organized as follows. In the next Section, we develop a theory for
1D junctions. We start with a discussion of basic equations and adopted
approximations, calculate the spectral characteristics of the junction using
Usadel equations, and finally obtain both the electric and the heat currents
through the junction. In Sec.~\ref{secplanar} we extend this theory to the case
of planar junctions. We discuss the results in Sec.~\ref{dis} and then consider
possible extension of our theory to the case of a double-barrier SINIS junction
in Sec.~\ref{sinis}. Finally, we summarize the results in Sec.~\ref{secconcl}.

\section{1D NN$'$IS junction model}\label{sec1D}

\subsection{Basic equations}

\begin{figure}[t]
\epsfxsize=7cm\epsffile{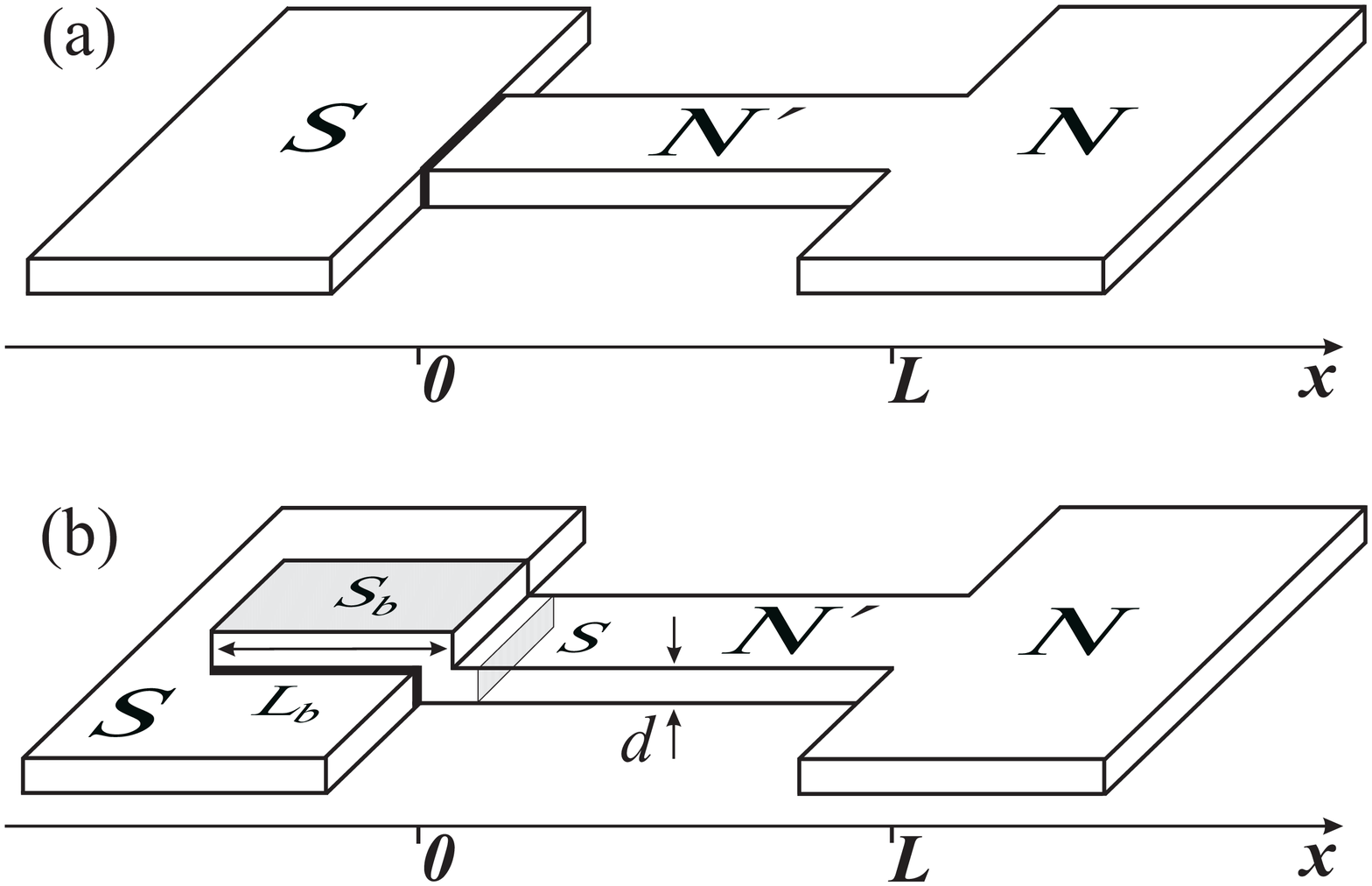} \vspace{-2mm}
\caption{One-dimensional (a) and planar (b) models of the NN$'$IS junction.
The insulating barrier is shown by thick black line.} \label{model} \vspace{-4mm}
\end{figure}

The model of the one-dimensional NN$'$IS junction is depicted in
Fig.~\ref{model}(a). It consists of a voltage-biased normal me\-tal reservoir
(N) and a normal metal wire (N$'$) of length $L$ connected to a superconducting
reservoir (S) through an insulator layer (I). We assume the NN$'$ interface to
be fully transparent.

In our theoretical analysis, we consider the diffusive limit, in which the
superconducting coherence length is given by expression $\xi _0 =
\sqrt{\mathcal{D}/2\Delta}$, where $\mathcal{D}$ is the diffusion coefficient
of the normal metal (we assume $\hbar = k_B = 1$) and the elastic scattering
length $\ell \ll \xi_0$. In this case, calculation of the electric and heat
currents requires solution of the one-dimensional Keldysh-Usadel
equations\cite{LOnoneq} (see also the review\cite{Belzig}) for the $4 \times 4$
matrix Keldysh-Green function $\check{G}(x, E)$ in the N$'$ lead,
\begin{align}\label{GK}
\bigl[\sigma_z E, \check{G}\bigr] = i \mathcal{D} \partial_x \check{J},
\quad \check{J} = \check{G}\partial_x \check{G}, \quad \check{G}^2 = 1.
\\
\check{G} = \begin{pmatrix} \hat{g}^R & \hat{G}^K \\
0 & \hat{g}^A
\end{pmatrix}, \quad \hat{G}^K = \hat{g}^R \hat{f} - \hat{f}
\hat{g}^A. \label{defG}
\end{align}
Here $\sigma_z$ is the Pauli matrix, $\partial_x \equiv \partial/\partial x$,
$\hat{g}^{R,A}$ are the $2 \times 2$ Nambu matrix retarded and advanced Green
functions, and $\hat{f} = f_+ + \sigma_z f_-$ is the matrix distribution
function (we use `check' for $4 \times 4$ and `hat' for $2 \times 2$ matrices).
In \Eqs{GK} we neglect the inelastic collision term, assuming the length $L$ of
the N$'$ lead to be smaller than the inelastic relaxation length.

Equations \eqref{GK} can be decomposed into the diffusion equations for the
Green functions,
\begin{equation}
[\sigma_z E, \hat{g}] = i \mathcal{D}\partial_x \hat{J}, \quad \hat{J} =
\hat{g}\partial_x \hat{g}, \quad \hat{g}^2 = 1,\label{GK_spectr}
\end{equation}
and the equation for the Keldysh component $\hat{G}^K$,
\begin{align}
[\sigma_z E, \hat{G}^K] = i \mathcal{D}\partial_x \hat{J}^K,\label{KineticG1}
\quad \hat{J}^K  &= \hat{g}^R
\partial_x\hat{G}^K + \hat{G}^K \partial_x \hat{g}^A.
\end{align}
Taking into account the normalization condition $\hat{g}^2 = 1$, we
parameterize the Green function by the complex spectral angle $\theta$,
\begin{equation}\label{Green}
\hat{g}(x,E)= \sigma_z \cosh\theta + i\sigma_y \sinh\theta.
\end{equation}

The electric and energy currents are related to the Keldysh component of the
matrix current $\check{J}$ as\cite{LOnoneq, Belzig, Vinokur, Golubov}
\begin{align}\label{I}
I &= \frac{g_N}{e} \int_0^{\infty} I_- \; dE, \quad Q = \frac{g_N}{e^2}
\int_0^{\infty} E I_+ \; dE,
\\
\label{Ipm} I_- &\equiv \frac{1}{4} \tr \sigma_z \hat{J}^K = D_- \partial_x
f_-, \quad I_+ \equiv \frac{1}{4} \tr \hat{J}^K = D_+ \partial_x f_+,
\end{align}
where $g_N$ is the normal conductance of the N$'$ lead per unit length, and
$D_\pm$ are dimensionless diffusion coefficients,
\begin{subequations}\label{Dpm}
\begin{align}
D_- & = (1/4) \tr ( 1 - \sigma_z \hat{g}^R \sigma_z \hat{g}^A ) = \cosh^2
\left( \re \theta \right),
\\
D_+ &= (1/4) \tr ( 1 - \hat{g}^R \hat{g}^A ) = \cos^2 ( \im \theta ).
\end{align}
\end{subequations}
Here we expressed the advanced  Green functions through the retarded ones
using the general relation $\hat{g}^A = - \sigma_z \hat{g}^{R \dagger}
\sigma_z$,\cite{LOnoneq} then omitted the superscript $R$. The quantity $I_+$
has the meaning of the spectral density of the net probability current of
electrons and holes, while $I_-$ represents the spectral density of the
electron-hole imbalance current responsible for the charge transfer (see the
discussion in Refs.~\onlinecite{Bezuglyi2000} and \onlinecite{Belzig}).

Calculation of the electric and energy currents in \Eqs{I} involves two steps:
first one has to solve the diffusion equations \eqref{GK_spectr} for the
spectral angle $\theta$, and then to solve the kinetic equations
\eqref{KineticG1} to find the distribution functions.

The expression for the heat current out of the normal metal reservoir (cooling
power) in a diffusive NIS structure was suggested by Bardas and Averin in
Ref.~\onlinecite{BA}. In contrast to the ballistic case (also discussed in
Ref.~\onlinecite{BA}), its spectral density contains several additional terms
which are odd in energy and therefore vanish upon integration over energy. We
propose another method for the derivation of the cooling power, which seems to
be physically clearer and does not involve the non-physical, odd-in-energy
terms. We define the heat generation in the reservoir through the work done by
the applied voltage on the nonequilibrium quasiparticles coming to this
reservoir, i.e., through the change of the kinetic energy $E_k$ of the
quasiparticles. We accept the definition $E_k^e=E-e\varphi(x)$ for the
electron-like and $E_k^h=E+e\varphi(x)$ for the hole-like quasiparticles, where
$\varphi(x)$ is the electric potential (note that the quantity $E$ is the total
quasiparticle energy which is conserved during passage across the junction, in
contrast to $E_k$). Along this line of reasoning, the heat generation in a
given reservoir can be defined as the kinetic energy flow to that reservoir,
\begin{equation} \label{I_k}
I_k(x) = \pm N_F S\int_{-\infty}^\infty [ E_k^e(x) I^e + E_k^h(x) I^h ]\,dE.
\end{equation}
We take $x=0$ and the minus sign for the left S reservoir, and $x=L$ and the
plus sign for the right N reservoir; $N_F$ is the electron density of states
per spin in the normal state, $S$ is the cross-sectional area of the junction,
and the quantities $I^e$ and $I^h$ are the electron and hole probability flow
densities, respectively. The expressions for $I^e$ and $I^h$ were found in
Ref.~\onlinecite{Bezuglyi2000} by introducing the following parametrization of
the matrix distribution function (see also Ref.~\onlinecite{Belzig}),
\begin{equation}
\hat{f} = 1 - 2\begin{pmatrix} n^e & 0 \\
0 & n^h
\end{pmatrix}, \quad n^{e,h}=\frac{1}{2}(1-f^{e,h}),\quad f^{e,h} = f_+ \pm f_-.
\nonumber
\end{equation}
The functions $n^e$ and $n^h$ have the meaning of the electron and hole
population numbers, respectively, and approach the Fermi distribution in the
reservoirs. Then the electron and hole probability currents are defined
as\cite{Bezuglyi2000}
\begin{align}
I^{e,h} &= (1/2)\mathcal{D}(I_+\pm I_-) \label{Ieh}
\\
&= -(1/2)\mathcal{D}[ (D_+ \pm D_-)\partial_x n^e + (D_+ \mp D_-)\partial_x
n^h ].\nonumber
\end{align}
In the N reservoir ($\theta =0$, $D_\pm=1$), the currents $I^{e,h}$ are
naturally related to the electron and hole diffusion flows,
$I^{e,h}=-\mathcal{D}\partial_x n^{e,h}$. Within the N$'$ lead each current
$I^{e,h}$ generally consists of a combination of both electron and hole
diffusion flows, which reflects the coherent mixing of electron and hole states
in the proximity region. Upon substitution of \Eq{Ieh} into \Eq{I_k}, using the
relation $g_N = 2e^2 N_F \mathcal{D} S$, we obtain the well-known equation for
the heat current out of the normal metal reservoir (cooling power),
\begin{equation}\label{heat}
P = -I_{k}(L) = - IV - Q.
\end{equation}
In the case of a NIN (Normal metal - Insulator - Normal metal) structure, the
heat generation in both reservoirs was found to be equal to
$IV/2$.\cite{Gurevich} For the NIS structure, it is the imbalance between the
kinetic energy flows to the N and S reservoirs that leads to the cooling
effect. The heat $P$ taken from the N electrode is then released in the S
reservoir, $I_k(0)=-Q=P+IV$, thus the full heat production in both reservoirs
is equal to the Joule heating, $I_k(0)+I_k(L)=IV$.

Now we discuss the boundary conditions. At $x = L$, we assume all functions
to be continuous, neglecting a spreading resistance of the transparent
NN$'$ interface: in the diffusive limit, this resistance is always small
compared to the resistance of the N$'$ wire.\cite{Nikolic} At the tunnel
barrier, $x=0$, the function $\check{G}$ and the matrix current $\check{J}$
at the normal (N$'$) and the superconducting (S) sides of the junction are
connected via the generalized boundary condition due to
Nazarov,\cite{Nazarov}
\begin{equation}\label{Nazarov}
\check{J}_{N'} = \frac{1}{2 g_N R_T} \int_0^1 \frac{\Gamma \rho(\Gamma) d\Gamma
[ \check{G}_S, \check{G}_{N'} ]}{1 + \frac{\Gamma}{4}( \{ \check{G}_S,
\check{G}_{N'} \} - 2 )},
\end{equation}
where $R_T$ is the barrier resistance and $\rho(\Gamma)$ is the distribution of
the transparencies of the conducting channels of the barrier $(\int_0^1 \Gamma
\rho(\Gamma) d\Gamma = 1)$. Assuming the absence of highly trans\-parent
channels with $\Gamma \sim 1$ and considering $\rho(\Gamma)$ to be localized
around a small value of $\Gamma \ll 1$ (tunnel limit), we can neglect the
anti-commutator term in \Eq{Nazarov}, thus arriving at the Kupriyanov-Lukichev
boundary condition\cite{KL} at $x=0$,
\begin{equation}
\check{J}_{N'} = (2 g_N R_T)^{-1} [\check{G}_S, \check{G}_{N'}].\label{KL}
\end{equation}
The boundary conditions for the functions $\hat{g}$ and $\hat{G}^K$ at the
tunnel barrier follow from \Eq{KL},
\begin{subequations} \label{Boundary2}
\begin{align}
\hat{J}_{N'}  &= (W / \xi_0)[\hat{g}_S, {\hat{g}}_{N'}], \label{BoundaryG}
\\
\hat{J}_{N'}^K  &= (W / \xi_0)[\check{G}_S, {\check{G}}_{N'}]^K.
 \label{BoundaryGK}
\end{align}
\end{subequations}
Here the tunneling parameter $W$ is defined as
\begin{equation}\label{W}
W = R(\xi_0)/2R_T = (3\xi_0/4\ell)\Gamma \gg \Gamma,
\end{equation}
where $R(\xi_0) = \xi_0 g_N^{-1}$ is the resistance of the N$'$ lead per length
$\xi_0$. It has been shown in Refs.~\onlinecite{Kupriyanov} and
\onlinecite{BBG} that it is this quantity, rather than the barrier transparency
$\Gamma$, that plays the role of the transparency parameter for diffusive
tunnel junctions. Below we consider the case $W \ll 1$, which corresponds to
the conventional tunneling limit.

The N and S electrodes are assumed to be equilibrium reservoirs with
unperturbed spectral characteristics and equilibrium quasiparticle
distributions,
\begin{align}
\theta_N &= 0, \quad f_{\pm N} = \frac{1}{2}\left( \tanh\frac{E + eV}{2T_N}
 \pm \tanh \frac{E - eV}{2T_N}  \right), \label{t0}
\\
\theta_S &= \arctanh \frac{\Delta}{E}, \quad f_{+S}  = \tanh \frac{E}{2T_S}
, \quad f_{-S} = 0,
\end{align}
where $T_N$ and $T_S$ are the temperatures of the N and S reservoirs,
respectively.

Using the parametrization in \Eq{Green}, we rewrite \Eq{GK_spectr} as the
Usadel equation\cite{Usadel} for the spectral angle $\theta(E, x)$,
\begin{equation}\label{Usadel}
i \mathcal{D} \partial^2_x \theta = 2E \sinh \theta.
\end{equation}
Here and below we omit the subscript N$'$ for the functions $f_\pm$ and $\theta$ in
the N$'$ lead. The boundary conditions for \Eq{Usadel} follow from \Eqs{t0} and
\eqref{BoundaryG},
\begin{subequations}
\begin{align}\label{Boundary0}
\theta\bigl|_{x = L} &= 0, \\
\partial_x \theta \bigl|_{x = 0} &= (2 W/\xi_0)
\sinh ( \theta_0 - \theta_S ),\label{BoundaryL}
\end{align}
\end{subequations}
where $\theta_0$ denotes the value of $\theta$ at $x=0$.

The kinetic equations for the functions $f_\pm$ follow from \Eq{KineticG1} and
have the form of conservation laws for the spectral currents $I_\pm$,
\begin{equation}\label{KineticG}
D_\pm\partial_x f_\pm = I_\pm = \mathrm{const}.
\end{equation}
The continuity of the distribution functions at the N$'$N interface implies the
conditions $f_\pm(E, L) = f_{\pm\, N}(E)$. The boundary conditions at the SN$'$
interface follow from \Eqs{BoundaryGK},\cite{Bezuglyi2000}
\begin{subequations}\label{KL_distr}
\begin{align}
g_N I_-(E) &= G^-_T(E) f_{-0}(E), \label{If-}
\\
g_N I_+(E) &=  G^+_T(E) [ f_{+0}(E) - f_{+S}(E) ],\label{If+}
\end{align}
\end{subequations}
where the subscript $0$ denotes the function values at $x=0$ and
\begin{align}
G^\pm_T(E) &= R_T^{-1}(N_S N_{N'} \mp M^{\pm}_S M^{\pm}_{N'}),\label{Gpm}
\\
N(E) &= \re( \cosh\theta ), \quad M^+(E) + i M^-(E) = \sinh \theta. \nonumber
\end{align}
The function $N(E)$ is the density of states (DOS) normalized to its value
$N_F$ in the normal state; the quantities $G_\pm$ can be interpreted as
spectral conductances of the tunnel barrier for the probability (+) and
electric (-) currents, respectively. At large energies, $|E| \gg \Delta$, when
$N(E)$ approaches unity and the condensate spectral functions $M^{\pm}(E)$ turn
to zero at both sides of the interface, the conductances $G^\pm_T(E)$ coincide
with the normal barrier conductance, $R_T^{-1}$. Within the subgap region $|E|
< \Delta$, $G^+_T(E)$ turn to zero, which reflects blocking of the probability
current due to full Andreev reflection.

In the superconducting reservoir, the density of states $N_S(E)$ and the
condensate spectral functions $M_S^\pm(E)$ read,
\begin{subequations}
\begin{align}
N_S(E) &= \frac{|E|\Theta(|E|-\Delta)}{\sqrt{E^2 - \Delta^2}},
\\
M_S^-(E) &= - \frac{\Delta \Theta(\Delta - |E|)}{\sqrt{\Delta^2 - E^2}},
\quad M_S^+(E) = \frac{\Delta \Theta(|E|-\Delta)}{\sqrt{E^2 - \Delta^2}},
\end{align}
\end{subequations}
where $\Theta(x)$ is the Heaviside step function.

\subsection{Solution of the Usadel and kinetic equations}

Generally, the solution of the Usadel equation for a N$'$ lead of finite length
can be found only numerically. However, in the case of a low-transparent tunnel
barrier, $W \ll 1$, the spectral angle is small, $\theta \ll 1$, for all
essential energies, which enables us to linearize \Eqs{Usadel} and
\eqref{BoundaryL},
\begin{subequations}\label{theta0L}
\begin{align}
i \mathcal{D} \partial^2_x \theta &= 2E \theta,\label{Usadel_dim}
\\
\partial_x \theta\bigl|_{x = 0} &= (2W/\xi_0) ( \theta_0\cosh\theta_S
- \sinh\theta_S).\label{thetaL'}
\end{align}
\end{subequations}
The analytical solution of these linearized equations,
\begin{align}\label{thetaz}
\theta(E, x) &= \theta_0(E) \frac{\sinh [k_N (L - x)/\xi_0]}{\sinh [k_N
L/\xi_0]}, \quad k_N = \sqrt{\frac{E}{i\Delta}},
\\
\theta_0(E) &= \frac{2W \sinh\theta_S}{k_N \coth(k_N L/\xi_0) + 2W \cosh
\theta_S},\label{thetaL}
\end{align}
was found to differ from the numerical solution of the exact, nonlinearized
Usadel equation by less than 1\% for reasonable values of $W \lesssim 10^{-2}$.
Note that in our approximation we keep a small term of the order of $W$ in the
denominator of \Eq{thetaL} which prevents divergence of $\theta_0$ at the gap
edge, $E=\Delta$, and thus provides a good agreement with the numerical
solution in the vicinity of this ``dangerous'' point.

The analytic solution of the kinetic equations \eqref{KineticG} with
corresponding boundary conditions \eqref{KL_distr} is
\begin{subequations}\label{f}
\begin{align}
&f_- = f_{-N} - \frac{f_{-N} R_N\alpha_-(x)}{R^-_T(E) + R_N\alpha_{-}(0)}
,\label{f_-}
\\
&f_+ = f_{+N} - \frac{(f_{+N} - f_{+S})R_N\alpha_+(x)}{R^+_T(E)  +
R_N\alpha_{+}(0)} ,\label{f_+}
\\
&\alpha_\pm(x) = \int_x^L \frac{dx'}{L}D_\pm^{-1}(E, x'),\nonumber
\end{align}
\end{subequations}
where $R^\pm_T(E) = \left[G^\pm_T(E)\right]^{-1}$ are spectral resistances of
the tunnel barrier,\cite{Bezuglyi2000} and $R_N = Lg_N^{-1}$ is the normal
resistance of the N$'$ lead. In \Eqs{f}, we used the relation
\begin{equation}\label{r1D}
{R_N}/{R_T} = 2W ({L}/{\xi_0}),
\end{equation}
following from the definition of the parameter $W$ in \Eq{W}. In a typical
experimental situation, the tunnel resistance dominates, $R_T \gg R_N$,
therefore the functions $f_\pm$ are always close to the equilibrium
distributions $f_{\pm N}$ in the N reservoir.

\subsection{Electric current} \label{elcur}

The electric current is given by the equation obtained by combining \Eqs{I},
\eqref{If-} and \eqref{f_-},
\begin{align}
I = \frac{1}{e} \int_0^{\infty} \frac{f_{-N}(E)}{R^-_T(E) + R_N^-(E)}\,dE ,
\label{II}\quad R_N^-(E) = R_N \alpha_{-}(0).
\end{align}
A similar result has been obtained for a NINIS structure in
Refs.~\onlinecite{VK} and \onlinecite{Bezuglyi2000}; it differs from \Eq{II} by
an additional tunnel resistance of the NIN interface in the denominator. In the
spirit of circuit theories for mesoscopic superconducting
structures,\cite{Nazarov,Bezuglyi2000} this equation can be interpreted as
``Ohm's law'' for the spectral current induced by the effective potential
$f_{-N}$ in the series of the tunnel resistance $R^-_T$ and the resistance
$R_N^-$ of the N$'$ lead renormalized by the proximity effect.

The current in \Eq{II} involves contributions of both the single-particle and
the two-particle (Andreev) currents. It is useful to discuss these two
components separately. To this end, we divide the total range of energy
integration into two regions, $E > \Delta$ and $0 < E < \Delta$, and take into
account that the superconducting DOS $N_S(E)=0$ at $0 < E < \Delta$ and the
spectral function $M_S^-(E)=0$ at $E > \Delta$,
\begin{align}
I = I_1 + I_A &= \frac{1}{e} \int_\Delta^\infty \frac{f_{-N}(E)}{R_T(N_S
N_{N'})^{-1} + R_N^-(E)}\,dE \nonumber
\\
 &+ \frac{1}{e} \int_0^\Delta
\frac{f_{-N}(E)}{R_T(M_S^- M_{N'}^-)^{-1} + R_N^-(E)}\,dE . \label{I1+IA}
\end{align}
The main contribution to the current $I_1$ comes from the processes of
single-particle tunneling. Besides, $I_1$ contains small proximity
corrections due to deviations of the DOS $N_{N'}$ and of the diffusion
coefficient $D_-$ in the N$'$ lead from their unperturbed values $N_N=D_N=1$.
Physically, these deviations are due to the partial Andreev reflection at the
energies above the superconducting gap and therefore rapidly decay as the
energy increases. Neglecting this small effect, we obtain the formula
\begin{align}
I_1= \frac{1}{e} \int_\Delta^\infty \frac{f_{-N}(E)}{R_T \,N_S^{-1} + R_N}\,dE
, \label{I1app}
\end{align}
which describes the single-particle current in the NN$'$IS structure. At the
subgap voltages, $eV < \Delta$, this current tends to zero exponentially at
small temperatures, $T_N \ll \Delta$. At large voltage, $eV \gg \Delta$, the
current $I_1$ approaches an Ohmic dependence with the deficit current arising
from the contribution of the N$'$ lead to the net junction resistance $R=R_T +
R_N$,
\begin{align}
I_1 &\approx \frac{V}{R} -I_{\textit{def}}, \quad I_{\textit{def}} \approx
\frac{r\Delta}{eR}\ln\frac{\sqrt{2}}{r}, \quad r =\frac{R_N}{R_T}. \label{r}
\end{align}
Finally, neglecting the small contribution $R_N$ to the junction resistance and
rewriting $f_{-N}$ in terms of the Fermi function of the N reservoir,
$n_N(E)=[1+\exp(E/T_N)]^{-1}$,
we arrive at the standard formula of the tunnel theory,\cite{Werthamer}
\begin{align}
I_1 = \frac{1}{eR_T}\int_{-\infty}^{\infty} N_S(E) [
n_N(E-eV)-n_N(E)]\,dE.\label{IT}
\end{align}

Within the same approximations, the Andreev current is reduced to the following
form,
\begin{align}
&I_A = \frac{2 W \Delta^2}{e R_T} \int_0^\Delta \frac{g_+ }{g_+^2 + ( g_- + 2 W E
)^2} \frac{f_{-N}(E)}{\sqrt{\Delta^2 - E^2}}\; dE.\label{IA-1D}
\\
&g_\pm(E) = \frac{\sinh \beta \pm \sin \beta} {\cosh \beta - \cos \beta}
\sqrt{\frac{E(\Delta^2 - E^2)}{2\Delta}}, \quad \beta =\sqrt{\frac{2
E}{\Delta}}\frac{L}{\xi_0}.\nonumber
\end{align}

At large voltage, $eV \gg \Delta$, the Andreev current approaches a constant
value $I_{exc} \approx ({\sqrt{2} W \Delta}/{e R}) \ln ({\sqrt{2}}/ {W})$
(excess current); for long junctions, $L \gg \xi_0$, it is much smaller than
the single-particle deficit current in \Eq{r}. Thus in this limit the net
electric current, $I_1 + I_A$, always exhibits a deficit current. The Andreev
current in NIS structures was first calculated microscopically by Hekking and
Nazarov \cite{Hekking} (see also Ref.~\onlinecite{flux}) and Vol\-kov {\it et
al}.\cite{VZK} Note that in our consideration we neglect possible pair-breaking
factors (like magnetic impurities) and damping of quasiparticles in the S
region due to inelastic interactions. For this reason, our results concerning
Andreev current may differ from that by Volkov {\em et al},
\cite{VZK,VolkovSIN} especially at small $eV$ comparable with corresponding
relaxation rates.

We would like to notice that the Andreev current does not depend on the N$'$
lead length $L$ as long as $L \gg \xi_0$. In this case, the magnitude of the
Andreev current at $eV \sim \Delta$ can be estimated from \Eq{IA-1D} as $I_A
\sim W\Delta/eR_T = \Delta R(\xi_0)/2e R_T^2$. This reproduces the result of
Hekking and Nazarov \cite{Hekking} and Volkov {\it et al}.\cite{VZK} As energy
decreases, the spectral density of the Andreev
current [the integrand in \Eq{IA-1D}] diverges as $E^{-1/2}$ until $E$ reaches
the small Thouless energy $E_{\textit{Th}} = \mathcal{D}/L^2$, which plays the
role of a cut-off factor. Such behavior of the Andreev current was first discovered
in Ref.~\onlinecite{Hekking} using the diagrammatic methods in the tunnel
Hamiltonian formalism. In the limit of a short junction, $L \ll \xi_0$, when
the proximity effect and the Andreev current are suppressed by the N reservoir,
we recover the result of Bardas and Averin,\cite{BA} $I_A \sim \Delta R(L)/2e
R_T^2 = \Delta R_N/2e R_T^2$.

\subsection{Energy current}

The energy current can be obtained upon combining \Eqs{I}, \eqref{If+}, and
\eqref{f_+},
\begin{align}
Q = \frac{1}{e^2} \int_0^{\infty} E\frac{f_{+N}(E)-f_{+S}(E)}{R^+_T(E) +
R_N^+(E)}\,dE , \quad R_N^+(E) = R_N \alpha_{+}(0). \label{Q}
\end{align}
This expression is quite similar to \Eq{II} for the electric current and has
the same physical interpretation: the spectral probability current flowing
through the series of the tunnel and normal resistances is determined by Ohm's
law for the effective potential difference $f_{+N}-f_{+S}$.

First, we note that the energy integration in \Eq{Q} is actually confined to
the interval $E>\Delta$ since the conductivity $G_T^+$ turns to zero (and,
correspondingly, $R_T^+ \to \infty$) at $0<E<\Delta$. Thus the Andreev energy
current $Q_A$ is identically zero; physically, this corresponds to the fact
that the quasiparticle probability current $I_+$ is completely blocked in the
subgap energy region due to full Andreev reflection.

Neglecting the proximity corrections to the spectral functions, i.e., assuming
$N_{N'} = D_+ = 1$ and $M^+_{N'} = 0$, we obtain a simplified form of the
single-particle energy current,
\begin{align}
Q_1= \frac{1}{e^2} \int_\Delta^\infty E\frac{f_{+N}(E)-f_{+S}(E)}{R_T
\,N_S^{-1} + R_N}\,dE. \label{Q1app}
\end{align}
Finally, omitting the contribution $R_N$ of the normal lead to the total
resistance and expressing $f_+$ in terms of the Fermi functions we arrive at
the standard form for the energy current,
\begin{equation}\label{Q_eq}
Q_1 = -\frac{1}{e^2R_T} \int_{-\infty}^{+\infty} N_S(E) E [ n_N(E-eV) -
n_S(E)]\,dE,
\end{equation}
where $n_S(E)=[1+\exp(E/T_S)]^{-1}$ is the Fermi function of the S reservoir.

\subsection{Heat current}

The heat current out of the normal metal reservoir (cooling power) can now be
obtained from the above expressions for the electric and the energy currents,
\Eqs{I1+IA} and \eqref{Q}, using \Eq{heat}. As follows from \Eq{heat}, the Andreev
heat current to the normal reservoir is nonzero giving a negative contribution
$P_A$ to the cooling power,
\begin{equation}\label{P_A}
P = P_1 +P_A, \quad P_A = - I_A V.
\end{equation}
From this equation we see that the heat current out of the normal metal is
affected by the Joule heating generated by the Andreev current $I_A$. This is due
to the fact that the Andreev current is  fully dissipated in the normal metal.

Using the tunnel model formula \eqref{IT} for the electric current and
\Eq{Q_eq} for the energy current, we arrive at the well-known form for the
cooling power,\cite{LPA}
\begin{equation}\label{P_eq}
P_1 = \frac{1}{e^2R_T} \int_{-\infty}^{+\infty} N_S(E) (E - eV) [ n_N(E-eV) -
n_S(E)]\,dE.
\end{equation}
This equation is widely used when fitting the experimental data on electron
cooling. Such an approach is valid as long as the Andreev contribution to the
electric current is negligibly small, i.e., at moderately high temperatures.
As noted above, at low temperatures, the single-particle processes are
exponentially suppressed in the subgap voltage region, where the effect of
Andreev current on electron cooling becomes essential and must be taken into
account.

\section{Planar NN$'$IS model}\label{secplanar}

In this section we present an extension of the approach developed above to the
more realistic case of a sandwich-type tunnel junction with a thin-film N$'$
lead as sketched in Fig.~\ref{model}(b). This situation is more complex;
however, it is possible to reduce this problem to the 1D case by formulating
effective boundary conditions at the junction, following a method suggested by
Volkov\cite{VolkovSIN} and Kupriyanov.\cite{Kupriyanov2D}

In the general three-dimensional case, the Keldysh-Usadel equations \eqref{GK}
and the boundary condition \Eq{KL} read
\begin{subequations}
\begin{align}
[\sigma_z E, \check{G}] &= i \mathcal{D} \nabla \check{\bf J}, \quad \check{\bf
J} = \check{G}\nabla\check{G},\label{GK3D}
\\
{\bf n} \; \check{\bf J}_{N'} &= ( 2 g_N R_T )^{-1} [ \check{G}_S,
\check{G}_{N'} ],\label{KL3D}
\end{align}
\end{subequations}
where $\bf n$ is a vector normal to the insulator layer. In \Eq{KL3D} all
functions are taken at the sides of the barrier.

We suppose the size of the planar junction $L_b$ to exceed the coherence
length, $L_b \gg \xi_0$, and the thickness of the N$'$ lead to be much smaller
than the coherence length, $d \ll \xi_0$. Then the function $\check{G}$ in the
left-hand side of \Eq{GK3D} is approximately constant within the normal metal
bank above the junction.\cite{Kupriyanov2D, VolkovSIN} Upon integration of this
equation over the volume of the normal metal bank, transforming the volume
integral in the right-hand side into a surface integral, and using the boundary
condition \Eq{KL3D} at the tunnel barrier, we obtain the effective boundary
condition for the 1D Keldysh-Usadel equations in the N$'$ lead,
\begin{equation}\label{Bound_V1}
S_b d [\sigma_z E, \check{G}_{N'}] = i \mathcal{D}\{ S \check{J}_0 - S_b
(W/\xi_0)[\check{G}_S, \check{G}_{N'}] \},
\end{equation}
where all functions are taken at the sides of the barrier. In \Eq{Bound_V1},
$S$ is the cross-section area of the N$'$ lead, $d$ is the lead thickness,
$S_b$ is the area of the junction [see Fig.~\ref{model}(b)], and $\check{J}_0 =
\check{G}_{N'}\partial_x \check{G}_{N'}$ is the value of the matrix current in
the N$'$ lead at the cross-section adjoining the junction (i.e., at $x=0$). A
similar result has been obtained in Ref.~\onlinecite{paper3} for the case of a
planar SIS junction. Equation \eqref{Bound_V1} can be rewritten as
\begin{align}\label{Bound_V2}
[\sigma_z E, \check{G}_{N'}] = 2 i \Delta \{ (\xi_0^2/L_b)\check{J}_0 -
\widetilde{W}[\check{G}_S, \check{G}_{N'}] \},
\end{align}
where
\begin{equation}\label{Wtilde}
\widetilde{W} = W(\xi_0/d)= ({3\xi_0^2}/{4\ell d}) \Gamma
\end{equation}
is the effective tunneling parameter. Note that for thin-film planar junctions
this parameter is much larger than the 1D tunneling parameter $W$ by the ratio
$\xi_0/d \gg 1$.

As long as $\xi_0 \ll L_b$ and $\xi_0 \check{J}_0 \sim W$, the first term in
the right-hand side of \Eq{Bound_V2} can be assumed to be the smallest one and
thus neglected. However, this is only true for the Green component of
\Eq{Bound_V2},
\begin{align}
[\sigma_z E, \hat{g}_{N'}] = 2 i \Delta \widetilde{W}
[\hat{g}_{N'}, \hat{g}_S], \label{Boundg}
\end{align}
whereas for the Keldysh component the diagonal part of the left-hand side
of \Eq{Bound_V2} turns to zero, and therefore the boundary condition for the
diagonal part of $\hat{J}_0^K$ reads
\begin{subequations}
\begin{align}
\hat{J}_0^K &= (W_f/\xi_0) [\check{G}_S, \check{G}_{N'}]^K,\label{Boundf}
\\
W_f &= W (L_b/d) = W (S_b/S) = \widetilde{W} (L_b/\xi_0) \gg
\widetilde{W}.\label{Wf}
\end{align}
\end{subequations}
The enhancement of the parameter $W_f$ with respect to $W$ reflects decrease of
the tunnel resistance $R_T$ compared to its value in the 1D case, due to
increase of the junction area from $S$ for the 1D geometry to $S_b$ for the
planar geometry (provided the barrier transparency is equal for both cases).

In terms of the spectral angle $\theta$ in the N$'$ lead, the boundary
condition \eqref{Boundg} has the form $k_N^2 \sinh \theta_0 = 2 \widetilde{W}
\sinh( \theta_S - \theta_0 )$
and can be solved explicitly for the boundary value $\theta_0$,
\begin{equation}\label{thetaL2D}
\theta_0 = \arctanh \frac{2 \widetilde{W} \sinh \theta_S}{k_N^2 + 2
\widetilde{W} \cosh \theta_S}.
\end{equation}
Equation \eqref{thetaL2D} results in a DOS minigap in the normal bank of the
junction. To first order in $\widetilde{W}$ this minigap is equal to
\begin{equation}
\Delta_g \approx 2 \widetilde{W} \Delta \ll \Delta,
\end{equation}
and the spectral angle $\theta_0$ is given by the BCS-like formula $\theta_0
\approx \arctanh (\Delta_g/E)$ at small energies, $E \sim \Delta_g$. The
spatial dependence of the spectral angle in the N$'$ lead obeys \Eq{thetaz}
with $\theta_0$ defined in \Eq{thetaL2D}.

We note that a similar result was found for short SINIS
junctions.\cite{VolkovSINIS,3T,Brinkman} The analogy between the NIS sandwich
and a short SINIS junction can be clearly seen from the mapping method, similar
to the one used in electrostatic problems. Indeed, at the top surface of the N
bank the boundary condition reads $\partial \theta /\partial {\bf n} = 0$. To
ensure this condition, we add a mirror image of the NIS sandwich to the top
surface of the N layer, thus arriving to the problem of a SINIS junction with a
normal metal interlayer of thickness $2d$.

The boundary condition \eqref{Boundf} for the distribution functions is similar
to \Eq{BoundaryGK} in the 1D case, with the substitution $W \rightarrow W_f$.
As follows from the definition of the parameter $W_f$ in \Eq{Wf}, the ratio of
the normal and tunnel resistances is similar to \Eq{r1D}, $R_N/R_T = 2 W_f
L/\xi_0$; therefore, the distribution functions in the planar geometry, being
expressed in terms of the spectral resistances, coincide with the result for
the 1D case, \Eqs{f}. As a result, equations \eqref{I1+IA} and \eqref{Q} for the
electric and energy currents hold their form for the planar geometry, however
with different tunnel resistances $R_T^\pm$.

We note that within the main approximation in $\W$, the spectral density of the
Andreev current is nonzero only inside the minigap, $E \leq \Delta_g$. In this
energy region, the spectral functions $M_S^-$ and $M_{N'}^-$ are approximately
equal to $-1$ and $-\Delta_g/(\Delta_g^2-E^2)^{1/2}$, respectively. Using
\Eq{I1+IA} and neglecting small contribution of the N$'$ lead to the net
resistance, we obtain a simple expression for the Andreev current at $eV \gg
\Delta_g$ in the planar NIS junction,
\begin{equation}
I_A = \frac{1}{eR_T}\tanh\frac{eV}{2T_N}\int_0^{\Delta_g}M_S^-M_{N'}^- dE =
\frac{\pi\Delta_g}{2 eR_T}\tanh\frac{eV}{2T_N}.
\end{equation}

\section{Results and Discussion} \label{dis}

\begin{figure}[t]
\epsfxsize=8.5cm\epsffile{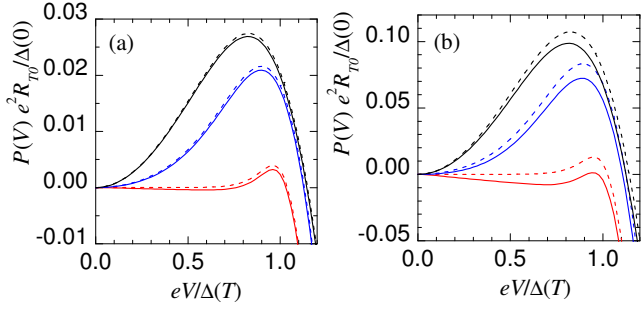} \vspace{-2mm}
\caption{(Color online) Cooling power versus bias voltage at $\W=0.5\cdot
10^{-3}$ (a) and $\W=2\cdot 10^{-3}$ (b) for different temperatures: $T = 0.1
T_c$ (red line), $T = 0.3 T_c$ (blue line), and $T = 0.5 T_c$ (black line).
Solid lines represent full cooling power, dashed lines were computed at
$I_A=0$.} \label{CoolingV} \vspace{-4mm}
\end{figure}

In our numerical calculations and analysis, we use exact expressions for the
electric, energy, and heat currents, \Eqs{I1+IA}, \eqref{Q}, and \eqref{heat},
taking into account both the proximity corrections and the resistance of the
N$'$ wire. Although these effects give small contributions to the energy and
electric currents separately, the cooling power $P$, being a relatively small
difference of the energy current and the Joule heat, is very sensitive to small
details of the charge and energy transport. We will focus on the case of a
planar junction which is the most adequate model of a real experimental setup.
In what follows, we assume the temperatures of the N and S reservoirs to be equal,
$T_N=T_S=T$.

We note that the tunneling parameters $W$ and $\widetilde{W}$, according to their
definition in \Eqs{W},\eqref{Wtilde}, are temperature-dependent, since the coherence
length $\xi_0$ increases with temperature as $\Delta^{-1/2}(T)$.
This variation is important at high enough temperatures
and it was taken into account in our calculation scheme, although at low
temperatures, $T \ll T_c$, when the cooling effect becomes apparent, one may
neglect the temperature dependence of $\Delta$. In what follows, we assume the
quantity $W$ and its effective value $\widetilde{W}$ in planar junctions to be
taken at $T=0$, allowing for their temperature dependence in \Eq{thetaL2D} for
the spectral angle by means of corresponding temperature-dependent factors. In
order to keep a common scale of the cooling power $P$ calculated for different
$T$ and $\W$, we normalize $P$ to the ratio $\Delta^2(0)/e^2 R_{T0}$, where
$R_{T0}$ is the junction resistance at a fixed value $\W=10^{-3}$ of the
tunneling parameter. Relying on typical sizes of the experimental samples, we
assume $L=L_b=10\xi_0$, where the coherence length $\xi_0$ is taken at $T=0$
(for Al-based film structures, its value is about 100 nm).

Now we proceed to the discussion of our results. The effect of the Joule heat
generated by the Andreev current $I_A$ on the cooling power is illustrated by
voltage dependencies $P(V)$ in Fig.~\ref{CoolingV}, where the solid curves
depict full cooling power, and the dashed curves were calculated at $I_A = 0$.
For a highly-resistive tunnel junction [$\W = 0.5\cdot 10^{-3}$,
Fig.~\ref{CoolingV}(a)], the heating effect due to Andreev current is
negligibly small. For smaller junction resistance [$\W=2\cdot 10^{-3} $,
Fig.~\ref{CoolingV}(b)], the heating effect essentially modifies the result; in
particular, at low enough temperature, $T=0.1T_c$, it makes $P(V)$ negative at
all voltages. This is due to the fact that for phase-coherent diffusive
proximity systems, the two-particle contribution to the subgap transport is
anomalously strong at low energies.\cite{Hekking, flux}
\begin{figure}[t]
\epsfxsize=8.6cm\epsffile{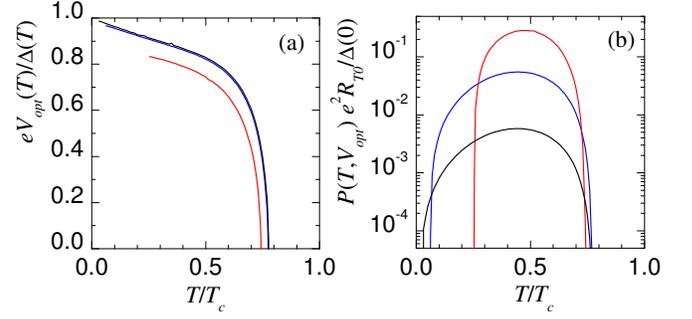} \vspace{-4mm}
\caption{(Color online) Temperature dependencies of the optimum bias
$V_{\textit{opt}}(T)$ (a) and of the cooling power $P(T)$ at optimum bias (b),
for different values of the tunneling parameter: $\W = 10^{-4}$ (black), $\W =
10^{-3}$ (blue), and $\W = 10^{-2}$ (red).} \label{CoolingT} \vspace{-4mm}
\end{figure}
\begin{figure}[t]
\epsfxsize=8.6cm\epsffile{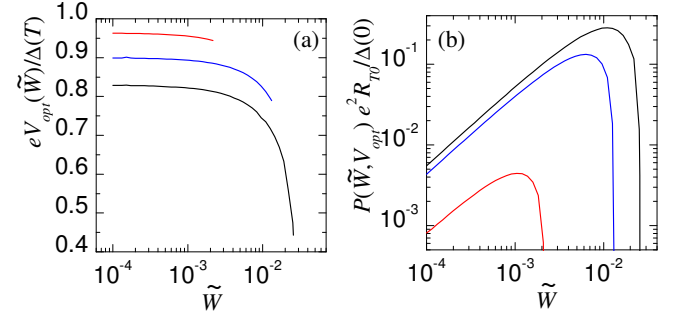} \vspace{-4mm}
\caption{(Color online) Dependencies of the optimum bias $V_{\textit{opt}}(\W)$
(a) and of the cooling power $P(\W)$ at optimum bias (b) on the tunneling
parameter, for different temperatures: $T = 0.1T_c$ (red), $T=0.3T_c$ (blue),
and $T=0.5T_c$ (black).} \label{CoolingW} \vspace{-4mm}
\end{figure}

As it is obvious from Fig.~\ref{CoolingV}, the cooling power approaches a
maximum at a certain optimal bias voltage $V_{\textit{opt}}$ which depends on
both the temperature and the tunneling parameter. It is interesting to note
that the dependence $V_{\textit{opt}}(T)$ is almost universal within a wide
range of the tunneling parameter, as shown in Fig.~\ref{CoolingT}(a). At $T
\gtrsim 0.75T_c$, the optimal bias formally turns to zero which means that at
these temperatures the cooling power becomes negative for all voltages.
Existence of the upper limiting temperature for the cooling effect is explained
by the increase in the number of thermally excited quasiparticles which produce
enhanced Joule heat. As the temperature decreases, the optimal bias rapidly
increases and approaches a value rather close to the energy gap $\Delta(T)$.
Simultaneously, the cooling power at optimal bias, $P[V_{\textit{opt}}(T)]$,
first increases and approaches a maximum at $T \approx (0.4\div0.5)T_c$, see
Fig.~\ref{CoolingT}(b). Then, at lower temperatures, the Joule heat due to
Andreev processes causes the cooling power to decrease. At a certain
temperature $T_{\textit{min}}$, the cooling power tends to zero, which defines
the lower limiting temperature for the cooling regime. As follows from
Fig.~\ref{CoolingT}(b), the temperature $T_{\textit{min}}$ increases with the
tunnel parameter, approaching $0.24T_c$ for $\W=10^{-2}$; this is because the
Andreev current and the associated Joule heat increase with the junction
transparency more rapidly than the single-particle cooling power. At
temperatures just above $T_{\textit{min}}$, the function $P(V)$ at small
applied voltage is negative; thus, the electron temperature is expected to
increase first with the bias due to Andreev current heating before it decreases
due to the single-particle cooling effect. This phenomenon has been observed in
experiments\cite{Sukumar1} at very low temperatures.

The dependencies of the optimal bias on the tunneling parameter $\W$ are
plotted in Fig.~\ref{CoolingW}(a). In accordance with the above-mentioned
universality of the curves $V_{\textit{opt}}(T)$ for different values of $\W$,
the dependence $V_{\textit{opt}}(\W)$ is rather weak at $\W \lesssim 10^{-2}$.
Within this region the cooling power at optimal bias,
$P[V_{\textit{opt}}(\W)]$, linearly increases with $\W$, as shown in
Fig.~\ref{CoolingW}(b), which is expected when single electron tunneling dominates.
For larger values of the tunnel parameter, the Andreev
current heating dominates over the single-particle cooling and leads to a rapid
decrease of the cooling power, which tends to zero at a certain onset point, as
seen from Fig.~\ref{CoolingW}(b). As the temperature decreases, the role of
Andreev processes becomes more important, therefore the onset shifts towards
smaller values of $\W$.

\section{Extension to SINIS junction}\label{sinis}

As noted in the Introduction, in most experiments the refrigerator is
arranged as a double-barrier SINIS junction,\cite{Sukumar1} where the S
electrodes are massive reservoirs, and N is a normal metal strip. Generally,
in such structures the charge and energy transport is due to multiple Andreev
reflections (MAR) of quasiparticles from the NS
interfaces.\cite{OTBK,Bezuglyi2000} During every passage across the junction,
the electrons and the retro-ref\-lec\-ted holes gain an energy $eV$, which
allows them eventually to overcome the energy gap and to escape into the S
reservoirs. This results in a strong quasiparticle nonequilibrium
characterized by intense electron heating within the subgap energy region,
which has been detected in the experiments.\cite{Pierre} From this point of
view the cooling effect observed in SINIS junction looks at a first glance
somewhat surprising.

However, the inelastic scattering processes impose strong limitations for the
existence of the MAR regime: in order to provide quasiparticle diffusion through
the whole MAR staircase, from $-\Delta$ to $\Delta$, the quasiparticle dwell
time in the N lead, $\tau_d$, must be smaller than the inelastic relaxation
time $\tau_\epsilon$ (for details, see Ref.~\onlinecite{Bezuglyi2000}). In
typical cooling experiments on SINIS junctions with low-transparent SN
interfaces,\cite{Sukumar1} the dwell time greatly exceeds $\tau_\epsilon$,
which prevents accumulation of the quasiparticle energy gains and thus destroys
the MAR regime. Correspondingly, the distribution functions in the N lead
become close to local-equilibrium ones,
\begin{align}
f_\pm(E,x) = \frac{1}{2}\left[ \tanh\frac{E + eV(x)}{2T_N}
 \pm \tanh \frac{E - eV(x)}{2T_N}  \right], \label{fSINIS}
\end{align}
where $V(x)$ is the voltage at the given point $x$. The variations in $V(x)$
are of the order of $V(R_N/R)$, i.e., negligibly small compared to the applied
voltage $V$ which mainly drops at the tunnel barriers. This implies that the
distribution functions are close to equilibrium functions in a normal
reservoir. In this case the SINIS junction behaves as two NIS junctions
connected in series through the equilibrium normal reservoir. As for the
spectral angle, in long junctions, $L \gg \xi_0$, it can be approximated by
the solution of the Usadel equation for a semi-in\-fi\-nite NIS structure;\cite{Vinokur}
simultaneously, this solution is also a good approximation to our solution for
a long NN$'$IS structure.

From this we conclude that our results can be applied to the description of
electron cooling in SINIS structures with large quasiparticle dwell times.
Similar modelling of a SINIS junction by a series of two NIS junctions has been
used in Ref.~\onlinecite{VZK} for the calculation of the differential conductance.

\section{Summary}\label{secconcl}

We have developed a quantitative theory of charge and heat transport in
one-dimensional and planar NN$'$IS tunnel junctions and studied the effect of
electron cooling in such structures. We extend the microscopic approach by
Bardas and Averin,\cite{BA} originally applied to constriction-type
junctions, to structures of arbitrary length and thin-film geometry used
in practice for microcooler fabrication. We found that the contribution of
two-particle (Andreev) current to the Joule heat generated in the normal
reservoir noticeably modifies the cooling effect, especially at low
temperatures and/or in rather transparent junctions. The interplay between
the Andreev current heating and the single-particle cooling, whose intensity
rapidly decreases with temperature, determines the lower limiting temperature
$T_{\textit{min}}$ for the cooling regime. When the transparency of the NIS
interface increases, the Andreev processes play a more essential role,
therefore the temperature $T_{\textit{min}}$ increases. At high
temperatures, the cooling regime is confined by the enhancement of the Joule heat
due to thermally excited quasiparticles; the maximal cooling temperature is
about $0.75T_c$, being almost independent of the junction resistance. As a
result, the cooling effect persists within a specific temperature interval
and approaches a maximum at the temperatures $(0.4\div0.5)T_c$.

We pay special attention to the analysis of the optimum bias voltage
$V_{\textit{opt}}(T,\W)$, at which the cooling power approaches a maximum for
given temperature $T$ and the tunneling parameter $\W$. We found that
$V_{\textit{opt}}$ exhibits a virtually universal temperature dependence for
different values of the tunneling parameter and approaches values close
to the energy gap as long as the temperature decreases. The cooling power at
optimum bias voltage first increases linearly with $\W$ until the Andreev
current heating abruptly suppresses the cooling regime.

We discussed the applicability of our results to the description of the cooling
effect in SINIS junctions. We show that such a double-barrier structure can be
modeled by a series of two independent NIS junctions, provided the
quasiparticle dwell time inside the junction greatly exceeds the inelastic
relaxation time. This condition, which is usually satisfied in cooling
experiments,\cite{Sukumar1} enables one to extend the theory presented here to the
case of the SINIS microcoolers.

From our considerations we conclude that the Andreev current is one of the most
serious factors of limitation of the electron cooling efficiency. In order to
reduce this factor, one should address materials in which the proximity effect
and, correspondingly, the Andreev current are strongly suppressed. A first
guess to such materials can be ferromagnets.\cite{Giazotto} A quantitative
analysis of the cooling effect in FIS junctions will be presented elsewhere.

\begin{acknowledgments}
The authors thank F.~Giazotto, A.A.~Golubov, T.T.~Heikkil\"{a}, J.P.~Pekola,
S.~Rajauria, F.~Taddei, and A.F.~Volkov for useful discussions. This work was
supported by Na\-no\-SciERA ``Na\-no\-fridge'' EU project.
\end{acknowledgments}

%%%%%%%%%%%%%%%%%%%%%%%%%%%%%%%%%%%%%%%%%%%%%%%%%%%%%%%%%%%%%%%%%%%%%%%%%%%

\end{document}